\documentstyle[aps,epsf]{revtex}    

\begin{document}
\twocolumn[\hsize\textwidth\columnwidth\hsize\csname 
@twocolumnfalse\endcsname                            

\author{A. Gammal$^{(a)}$, T. Frederico$^{(b)}$ and L. Tomio$^{(a)}$}
\address{$^{(a)}$
Instituto de F\'\i sica Te\'orica, Universidade Estadual Paulista, \\
01405-900 S\~{a}o Paulo, Brazil \\
$^{(b)}$Departamento de F\'{\i}sica, Instituto Tecnol\'ogico da
Aeron\'autica, \\
Centro T\'ecnico Aeroespacial, \\
12228-900 S\~ao Jos\'e dos Campos,
SP, Brazil }

\title{
Improved numerical approach for time-independent Gross-Pitaevskii 
nonlinear Schr\"odinger equation.
}
\maketitle

\begin{abstract}
In the present work, we improve a numerical method, developed to solve the
Gross-Pitaevkii nonlinear Schr\"{o}dinger equation. 
A particular scaling is used in the equation, which permits to evaluate
the wave-function normalization after the numerical solution.
We have a two point boundary value problem, where the
second point is taken at infinity.
The differential equation is solved using the shooting method and
Runge-Kutta integration method, requiring that the asymptotic constants, 
for the function and its derivative, are equal for large distances. 
In order to obtain fast convergence, the secant method is used. 
\vskip 0.3cm 
{PACS: 71.10.+x ,
02.60.Lj,
11.10.Lm,
03.75.Fi}
\end{abstract}
\vskip 0.5cm ]                              

Precise and fast numerical solutions to non-linear equations have 
become considerable important in computational physics.
So, the numerical procedure are relevant also to be described,
when treating such problems, considering the computational time 
consuming. \ 
In the present work, we pay attention specially to this
problem, proposing an alternative approach to a method recently
described in Ref.~\cite{edw95}, which was used to solve the 
Gross-Pitaevkii~\cite{gin} nonlinear Schr\"{o}dinger equation (NLSE) for
trapped neutral atoms, with positive two-body scattering lengths.  
In Ref.~\cite{rup95}, the NLSE treated in Ref.~\cite{edw95} was extended
to a time-dependent one, for both positive and negative two-body
scattering lengths, where the Crank-Nicolson algorithm
(appropriate for time evolution) was considered.
\ This approach, however, has the disadvantage that can only reach the 
stable solutions. 
In case one needs to add other non-linear terms (of higher order) 
in the original equation~\cite{col98}, it is not feasible
to reach another region of stable solutions if in between there is an 
unstable region. This implies that it should be appropriate to 
combine a static method (as the one used in Ref.\cite{edw95}) with the
method used in Ref.\cite{rup95} when we are interested in obtain all
the stable and unstable solutions and also the corresponding 
time evolutions. 
So, in this perspective, any improvement of the method considered
in Ref.~\cite{edw95} would be relevant.  

In the following, we briefly describe the physics related to 
the NLSE considered in \cite{edw95}, and the numerical procedure 
used to solve it. \ Then we present an alternative approach, which can 
considerably reduce the time to search for the solutions and the
normalizations. 

The nonlinear Schr\"{o}dinger equation, which describes the condensed
wave-function in the mean-field approximation can be
written as~\cite{edw95} 
\begin{equation}
\left[ -\frac{\hbar ^{2}}{2m}{\bf \nabla }^{2}+\frac{m}{2}\omega ^{2}r^{2}-
\frac{4\pi \hbar ^{2}\ |a|}{m}|\Psi (\vec{r})|^{2}\right] \Psi (\vec{r})\ =\
\mu \Psi (\vec{r})\ ,  \label{sch}
\end{equation}
where $m$ is the mass of a single atom, $\omega$ the angular frequency
of the trap, $\mu $ the chemical potential and $a$ the scattering
length. In the present approach, as we are more concerned with numerical 
aspects, for convenience we treat only cases with negative
scattering lengths ($a<0$)~\footnote{For the numerical
considerations, there is no restrictions about the sign of the scattering 
length, as the solutions with $a>0$ are equally accessible using the same
procedure and changing the sign of the non-linear term in 
Eq.~(\ref{sch}).}. \ Later we also consider the inclusion of a three body
interaction term.

The chemical potential $\mu$ is fixed by the number $N$ of atoms
in the condensed state, which is given by the normalization condition: 
\begin{equation}
\int d^{3}r|\Psi (\vec{r})|^{2}\ =\ N\ .  \label{norm1}
\end{equation}
In Refs. \cite{edw95} and \cite{rup95} the NLSE for Bose-Einstein 
condensates, as given in Eq.~(\ref{sch}), was solved numerically.
In Ref.\cite{edw95}, the shooting and the Runge-Kutta 
methods~\cite{pr92,qui87} 
were combined. For a given normalization parameter its was solved the
corresponding dimensionless equation.
The asymptotic form of the wave-function was renormalized to be equal to
the numerical wave-function for a sufficiently large distance.
The wave-function normalization parameter was
increased until the Wronskian of the asymptotic behavior of
the numerical and the analytic function changes sign. 

Next, we present in detail the numerical approach we have used, 
in order to show up the similarities and subtle differences between
this approach and the one of Ref.~\cite{edw95}
As we suggest from our experience, such subtle differences in the
numerical procedures will reduce considerably the time to search for
the solutions. 
We first rewrite Eq.(\ref{sch}) in dimensionless units, in order to become
apparent the physical scales contained in Eq.(\ref{sch}). \ By rescaling
Eq.(\ref{sch}) for the s-wave solution, we obtain~\cite{col98} 
\begin{equation}
\left[ -\frac{d^{2}}{dx^{2}}+\frac{1}{4}x^{2}-\frac{|\Phi (x)|^{2}}{x^{2}}
\right] \Phi (x)\ =\ \beta \Phi (x)\ ,  \label{schd}
\end{equation}
where $x\equiv \sqrt{{2m\omega }/{\hbar }}\ r$, $\Phi (x)\equiv 
\sqrt{8\pi |a|N}\;r\Psi (\vec{r})$ and $\beta \equiv {\mu }/{\hbar
\omega }\leq 3/2.$  The normalization for $\Phi (x)$, obtained from 
Eq.\ref{norm1},
defines a real number $n$ (given as $|C_{nl}^{3D}|$ in
Ref.~\cite{edw95}) related to the number of atoms $N$: 
\begin{equation}
\int_{0}^{\infty }dx|\Phi (x)|^{2}\ =n,\;\;\;{\text{where}}\;\;\;n\equiv
N(8\pi |a|)\sqrt{\frac{2m\omega }{\hbar }}\ .  \label{norm2}
\end{equation}

We would like to emphasize that, by using this scaling procedure, the
numerical solutions for the equation are free of any normalization
constraint, or other parameter dependence.
The parameter $N$, related to the number of particles, was removed from
the differential equation and its not necessary to check
Eq.~(\ref{norm1}) or (\ref{norm2}) at all steps of the calculation.
The normalization is found {\it a posteriori}, using Eq.~(\ref{norm2}).

The boundary conditions for Eq.~(\ref{schd}) are given
as~\cite{edw95} 
\begin{eqnarray}
&&\Phi (0)=0 \;\;\;\; {\rm and}\;\;\;\;
\Phi (x)|_{x\rightarrow \infty } 
\rightarrow \Phi_{asym}(x) \nonumber \\
&&
\Phi_{asym}(x)
\equiv C\exp{\left[-\frac{x^{2}}{4}+
\left(\beta -\frac 12\right)\ln (x)\right]} ,
\end{eqnarray}
where $C$ is a constant to be determined.  By using
Runge-Kutta method and starting with the given $\Phi (0),$ the problem
is reduced to determine the value of the corresponding derivative, 
$\Phi^{\prime }(0)$, which satisfies the asymptotic 
condition at infinity. \
So, we are tempted to shoot~\cite{pr92,qui87} many values for $\Phi
^{\prime }(0)$ until we obtain numerically a constant for large distances.
At a certain $x_{max}$ we expect a constant, given by 
\begin{equation}
C_{\Phi }\equiv \left.{\Phi _{num}(x)}\;{\exp\left[\frac{x^2}{4}-
\left(\beta -\frac 12\right)\ln(x)\right]}\right .
\label{cefi1}
\end{equation}
This process is very laborious and difficult to reach precise
solutions due to the problem of verifying, for some large
$x$, when $C_\Phi$ is constant, within the required numerical
precision.
The way to overcome these difficulties is to consider
the asymptotic derivative of $\Phi(x)$, that is 
\begin{eqnarray}
&&\Phi_{asym}^\prime(x) = \nonumber \\
&&C\left[-\frac x2+\left(\beta -\frac
12\right)\frac{1}{x}
\right]
\exp\left[{-\frac {x^2}{4}+\left(\beta -\frac 12\right)\ln (x)}\right]
\end{eqnarray}
and also determine (numerically) the expression 
\begin{eqnarray}
&&C_{\Phi ^{\prime }}\equiv 
\Phi _{num}^{\prime }(x) \times \nonumber \\ &&
\times
{\left[-\frac x2 +\left(\beta  -\frac 12\right)\frac{1}{x}\right]^{-1}
\exp{\left[\frac{x^2}{4}-\left(\beta -\frac 12\right)\ln
(x)\right]}}
\label{cefi2} ,
\end{eqnarray}
with $x_{max}$ such that both (\ref{cefi2}) and (\ref{cefi1}) are
constants. \  When we are using the correct value of $\Phi ^{\prime}(0)$
we also should obtain $C_{\Phi }=C_{\Phi ^{\prime }}=C$ for 
a large enough $x=x_{max}$. \ 

An useful remark we can make is that when we overestimate the value of
$\Phi^{\prime }(0)$, $C_{\Phi }$ increases and $C_{\Phi ^{\prime }}$
decreases. \ The inverse happens when we subestimate $\Phi ^{\prime }(0).$
So, this condition is valuable to  
tune $\Phi ^{\prime }(0)$. \ It corresponds to {\it solve the
equation}
\begin{equation}
C_{\Phi }-C_{\Phi ^{\prime }}=0,  \label{cefi}
\end{equation}
{\em having} $\Phi ^{\prime }(0)$ {\em as the unknown variable}.
Substituting  (\ref{cefi1}) and (\ref{cefi2}) \ in (\ref{cefi}) we recover
the expression for the Wronskian $W(\Phi _{num}(x_{\max }),\Phi
_{asym}(x_{\max }))=0$ stated in  \cite{edw95}. 
Eq.~(\ref{cefi}) can be solved by secant method~\cite{qui87}. 
So, we begin with an approximate solution for $\Phi^{\prime }(0)$, as an
input to the secant method, where \ $C_{\Phi }$ and $C_{\Phi
^{\prime }}$ are evaluated by the Runge-Kutta method. We should
emphasize that, to succeed with such method the original guess for $\Phi
^{\prime }(0)$ should be not far from the correct value, otherwise the
method can conduct to the trivial solution $\Phi (x)=0$ or to overflows.
In our procedure, for a fixed $\beta$, $x_{max}$ was first estimated 
to be equal to 4.2 and $\Phi^\prime(0)$ was used as an initial trial 
to extend $x_{max}$ to 5.6 and subsequently to 7.0. 
Once we find a solution for $\Phi^{\prime }(0)$, for a given $\beta $,
we  decrease $\beta $ slightly by $\Delta \beta $ using the previous 
$\Phi ^{\prime }(0)$ to find the new $\Phi ^{\prime }(0)$. 
This process allows us to ``walk'' along $\beta $ values, obtaining the
corresponding solutions and results for $n$. \ 

Although the secant method can become unstable under certain conditions,
in this case it will not occur, as we explain in the following.
We found that the secant method is appropriate, as we can be as near as
desired to the solution, starting with a given analytical solution of
the corresponding linear Schr\"odinger equation.
So, we  just need to implement an automatic algorithm routine to
make slow variations of $\beta$ and the corresponding slow shift (from the
initially zero) of $\Phi^\prime(0)$, in order to satisfy the corresponding
non-linear equation.
\ In this way, we are always near the solution, such that the secant method 
can be applied.
We think the same procedure can be generally applied for solitonic 
equations. The algorithm of slow variation of $\beta$ (deformation algorithm) 
does not need the estimation for the derivative of the wave-function at $x=0$,
as given in Eq.(3.7) of Ref.~\cite{edw95}) for every solution, except for the
first one were we take it near the harmonic oscillator solution.
We understand that the analytical approximation given in Ref.~\cite{edw95} to
estimate the derivative at $x=0$ is not the most convenient in the present
case. Considering that in general such equations are highly non-linear, the
initial guess for the derivative of $\Phi$ at $x=0$ can easily cause overflow
when determining the asymptotic constants (Wronskian) at large distances. \
Our initial guess  can be very close to the harmonic oscillator solution,
which
corresponds to $\Phi^\prime(0)=0$, avoiding possible overflows for 
sufficient large distances.
In our numerical approach,
considering that $\beta =1.5$ is a trivial solution of the linear harmonic
oscillator, we started with $\beta =1.4$, trial $\Phi^\prime(0)=0.6$ 
and $\Delta \beta =0.02.$
For each $\beta $ it was necessary $4\sim 6$ iterations in the secant
method~\cite{qui87}, for each $x_{max}$.

Our results, for several values of $\beta$, are partially listed in 
Table I. \ The solutions with $\beta\le 0.4$ are unstable and not 
shown in Ref.~\cite{rup95}. 
However, the solutions with $\beta\ge 0.4$ well agree with their results.
As one can observe in Table I, the method also can reach solutions with
negative chemical potentials ($\beta< 0$).
A numerical stability check, which can be done by evolving the static
solutions, can easily be followed by using a time dependent method, as the 
Crank-Nicolson method~\cite{rup95,qui87}.

In Fig. 1 we also show three plots for the chemical potential, $\beta$,
as a function of $n$, in case of zero angular momentum. \ The three plots
shown correspond to the lower radial states ($n_r = 0, 2, 4$) of
Eq.~(\ref{schd}).
The plot labeled with $n_r=0$ corresponds to Table I.
In the limit of the harmonic oscillator solution, where $n=0$ and the 
equation is linear, we obtain the usual known solutions.

In Fig.2 we have another example of the application of the method 
described here. In this case, we consider the addition of another 
non-linear term inside the square brackets of Eq. (\ref{schd}), given
by
\begin{equation}
g_{3}\frac{|\Phi (x)|^{4}}{x^{4}} ,
\end{equation}
which can be directly related to the three-body effects, where
$g_3$ is the non-dimensional strength of the corresponding 
three-body interaction.
 \ The physical consequences of the addition of such a term in the NLSE is
discussed in both references given in \cite{col98}. 

\begin{table}
\caption{Numerical solutions for the NLSE including two-body
interaction, when the two-body scattering $a$ is negative.
We consider that at $x_{\max }=7.0$ we have achieved the 
asymptotic limit.}
\begin{tabular}{cccc}
$\beta $ & $\phi ^{\prime }(0)$ & $C$ & $n$ \\ \hline\hline
1.5 & 0         &     0 & 0      \\
1.4 & 0.5448721 & 0.535 & 0.3310 \\
1.2 & 0.9939222 & 0.929 & 0.7854 \\ 
1.0 & 1.3567267 & 1.187 & 1.2282 \\ 
0.8 & 1.7022822 & 1.374 & 1.4607 \\ 
0.6 & 2.0495486 & 1.510 & 1.5839 \\ 
0.4 & 2.4045809 & 1.608 & 1.6254 \\ 
0.2 & 2.5851166 & 1.648 & 1.6237 \\ 
0.0 & 3.1340461 & 1.741 & 1.5632 \\
-1.0& 4.8924036 & 2.110 & 1.2234 \\ 
-2.0& 6.3914678 & 2.995 & 0.9843 \\
\end{tabular}
\end{table}

\begin{figure}  
\setlength{\epsfxsize}{0.9\hsize} \centerline{\epsfbox
{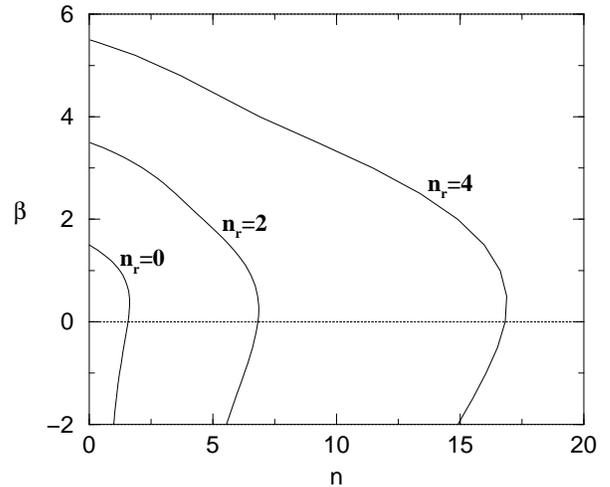}}

{\caption{
We show the chemical potential, $\beta$, as a function of 
$n$, which is related to the number of particles by the Eq.(\ref{norm2}).
The three plots shown correspond to the lower radial states ($n_r$) 
of Eq.~(\ref{schd}) (with zero angular momentum), such that in the limit of
the harmonic oscillator solution, where $n=0$, we have the usual 
known solutions.
The plot labeled with $n_r=0$ corresponds to Table I.}}
\end{figure}

\begin{figure}  
\setlength{\epsfxsize}{0.9\hsize} \centerline 
{\epsfbox{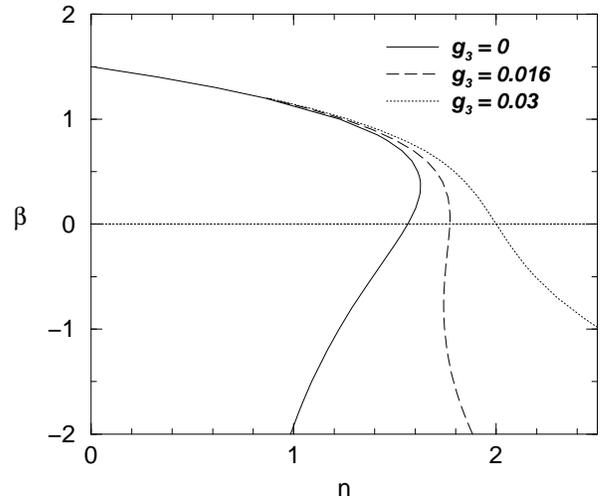}}

{\caption{
We show the chemical potential, $\beta$, as a function of 
$n$, in case we consider three-body effects.
The three plots shown correspond to the radial state solutions for
$g_3 =0$ (no quintic term), $g_3 =$ 0.016 and $g_3 =$ 0.03. 
The plot labeled with $g_3=0$ corresponds to Table I.
}}
\end{figure}

To finalize, we have presented in detail an improvement to
a numerical procedure used to solve a non-linear differential
equation, which is commonly applied for Bose-condensed states. In 
our example, we solve the Gross-Pitaevskii equation with an
attractive two-body and a repulsive three-body interaction. 
We should note that, by using a simplified scaling procedure [given in
Eqs.(\ref{schd}) and (\ref{norm2})], the numerical solutions for the
equation are free of any normalization constraint, or other parameter
dependence. 
The parameter $N$, related to the number of particles, was removed from
the differential equation and it is not necessary to obtain
the normalization at each step of the calculation.
Eq.~(\ref{norm2}) gives the normalization {\it a
posteriori}.
So, by using the above scaling procedure, it emegers the main difference
between the present method and the one given in Ref.~\cite{edw95}, 
when looking for the solutions of the NLSE: 
\begin{itemize}
\item[-] in our approach, we searched for the 
derivative of the wave-function at $x=0$ till the
asymptotic constants match (when the Wronskian vanishes), 
and the normalization is given at the end;
\item[-] in \cite{edw95} the normalization parameter $A$ is
incremented till the sign of the Wronskian is reached. \ For the 
final renormalization they also use other intermediate parameters, as 
$A_0$, $N_0$, $A_1$, $N_1$. 
\end{itemize}
As we are not restricted by the normalization, our approach is
a clear improvement to the method given in Ref.~\cite{edw95}, 
particularly when considering the simplification and the transparency in 
the normalization procedure. \ Such advantage can be further explored
when more involved calculation are presented, as in Ref.~[7]
where collective excitations are evaluated.
\ A different scaling removes the normalization constraint
and allows one to obtain it {\it a posteriori},
after the numerical solutions are achieved for the eigenvalues. 

In our numerical procedure, we employ the shooting method on Runge-Kutta
integration, matching the
asymptotic constants for the wave-function and for the corresponding 
derivative. \ This procedure is shown to be equivalent
to make the Wronskian vanish at such large distances. \ In order to obtain 
a faster convergence to the solution, we also included the secant method.
The numerical optimization to the method employed in Ref.\cite{edw95}, 
described here, is not restricted to the NLSE we have used. It can be used 
quite generically for second order solitonic differential equations whose 
solutions asymptotically vanish at large distances.\
We consider particularly relevant an optimization of the method of 
Ref.~\cite{edw95} in the perspective of looking for solutions of 
differential equations with higher order non-linear terms; and also 
in the perspective of combining such method (appropriate for static 
solutions) with a time-dependent one.

{\bf Acknowledgments}

This work was partially supported by Funda\c {c}\~{a}o de Amparo \`{a}
Pesquisa do Estado de S\~{a}o Paulo and Conselho Nacional de Desenvolvimento
Cient\'{\i}fico e Tecnol\'{o}gico.

\end{document}